\documentclass[11pt,twoside]{./atmp}

\usepackage{amsmath,amssymb}
\usepackage{graphicx}
\usepackage[all]{xy}
\usepackage{color}
\usepackage{mathptmx}
\usepackage{latexsym}

\begin{document}

\title[Homogeneous cosmologies in scalar tensor theory]
{Homogeneous cosmologies in scalar tensor theory}


\author[M. Ulu Do\~{g}ru and D. Baykal]{Melis Ulu Do\~{g}ru$^{a}$         \and
        Derya Baykal$^{b}$}

\address{$^{a}$ Department of Physics, Art and
Science Faculty, \c{C}anakkale Onsekiz Mart University \\
\c{C}anakkale, 17020, Turkey \\ $^{b} $Institute for Natural and Applied Sciences, \c{C}anakkale Onsekiz Mart University \\
\c{C}anakkale, 17020, Turkey}  
\addressemail{melisulu@comu.edu.tr}

\begin{abstract}
In this study, FRW-cosmologies with some matter groups such as monopole-domain wall, monopole-Chaplygin gas and monopole-strange quark matter in the scalar theory of gravitation based on Lyra geometry are investigated. We expand two exact models as static case and
time-depended case for each matter groups in order to solve field equations in the
scalar theory. For each matter groups, the solutions are introduced as the models of
expanding universe, exponentially. Hubble parameters
in the case of $k=0,-1,1$ are obtained for these models. Furthermore, we realize
interesting result which the well-known relation between scalar
theory based on Lyra geometry and Einstein's theory is an
incomplete idea. In opposition to the well accepted idea in the
literature, we suggest that Einstein's theory with no
cosmological constant is equivalent of scalar theory based on Lyra
geometry with zero displacement vector, completely. If
the components of displacement vector in the scalar theory are any constant functions, the
scalar theory couldn't correspond to Einstein's theory, identically.
Even if the components of displacement vector in the scalar theory are the constant,
the field equations and their solutions contain the Einstein's field
equations and their solutions, but they are variously more general
than Einstein's theory of gravitation. So, coefficients of constant displacement vector don't play the role of cosmological constant in keeping with the Einstein's theory. Finally, the results have
been discussed.
\end{abstract}

\maketitle

\section{Introduction}

The fundamentals of gravitation theories have been predicated on the
Newtonian Theory, early on. But, at the
end of 19th century, it was found that the velocity of light has finite value by well-known experiment carried out by Michelson and Morley. Consequently, the Lorentzian transformations invalidate the Galilean transformations. For the many reason, the new gravitation theories with Lorentz invariance was needed to arise. Suggested theories with Lorentz invariance can be classified as
Poincare-type, scalar-type, vectorial-type and tensorial-type
gravitation theories. In addition to the available theories at present,
adopted gravitation theories are called as Einstein's theory,
Teleparallel gravity, Brans-Dicke theory, modified gravity and
Lyra geometry. Both of them, Lyra \cite{lyra} and Brans-Dicke
\cite{brans}, are the alternative scalar-tensor theories
\cite{rahgos}. Indeed, an alternative scalar-tensor theory of
gravitation was propounded by Weyl \cite{weyl} in 1918 via
associated the electro-dynamical and gravitational states of a
space-time, firstly. But geometry of Weyl's theory was not valid
and it was not useful since it was depended on non-integrability
of length transfer \cite{pradplane}. Lyra evolved a new theory,
composed by modification of Riemannian geometry and based on Weyl's
geometry \cite{rahmalgos}. In Lyra geometry, the concept of scalar
curvature is defined as opposite of Weyl's theory \cite{pradcha}.

In Lyra geometry, $n-$dimensional space-time $(M, \phi , g_{ik})$
is contained a smooth manifold, a smooth scalar field and
connection, on condition that $M$, $\phi$ and $\Gamma$ are
introduced manifold, the gauge function and Lyra connection,
respectively \cite{kitap}. The coefficients of Lyra connection are
given by

\begin{equation}  \label{eq1}
\Gamma^{c}_{ab}=\frac{1}{\phi}\{\begin{array}{c}
           c \\
           ab
         \end{array}
\}+\frac{s+1}{\phi^2}g^{cd}(g_{bd}\partial_a \phi-g_{ab}\partial_d \phi)
\end{equation}
where $\{\begin{array}{c}
           c \\
           ab
         \end{array}
\}$ is the second kind of Christoffel symbols, $s$ is a constant. Also, torsion is introduced by

\begin{equation}  \label{eq2}
T_{ab}^{c}=\frac{s}{\phi^2}(\delta_{b}^{c}\nabla_a
\phi-\delta_{a}^{c}\nabla_b \phi).
\end{equation}
Curvature tensor in Lyra geometry is given by \cite{kitap}

\begin{equation}  \label{eq3}
K_{dab}^{c}=\frac{1}{\phi^2}[\partial_a (\phi\Gamma_{db}^{c})-\partial_d (\phi\Gamma_{ab}^{c})+\Gamma_{db}^{e}\Gamma_{ea}^{c}-\Gamma_{ab}^{e}\Gamma_{ed}^{c}]
\end{equation}
and also, contractions of curvature tensor are formed
$K_{ac}=K_{abc}^{b}$, $K=g^{ab}K_{ab}$, similar to Riemannian
geometry. Finally, scalar curvature is defined by the following
form:

\begin{equation}  \label{eq4}
K=\frac{R}{\phi^2}+\frac{2(s+1)}{\phi^3}(1-n)\Box\phi+\frac{1}{\phi^4}[(s+1)^2(3n-n^2-2)-2(s+1)(2-n)]\nabla^{c}\phi\nabla_{c}\phi
\end{equation}
where $\Box$ signs D'alambertian operator \cite{kitap}. The action
in scalar theory based on Lyra geometry is given by

\begin{equation}  \label{eq5}
S=\int d^4 x\sqrt{-g}(\phi^2 R-4\omega g^{cd}\nabla_{c}\phi\nabla_{d}\phi)
\end{equation}
where $\omega=\frac{3(s^2-1)}{2}$.

\begin{equation}  \label{eq6}
R_{ij}-\frac{1}{2}g_{ij}R+\frac{3}{2}\phi_{i}\phi_{j}-\frac{3}{4}g_{ij}\phi_{k}\phi^{k}=-8ðGT_{ij}
\end{equation}
where $\phi_{i}$ is the displacement vector.

Sen \cite{sen}, Sen and Dunn \cite{sendunn} reproduced the Lyra geometry. Halford
\cite{halford}, suggested that cosmological constant in the
Einstein's theory corresponds the constant displacement vector
field $\phi$ in scalar theory of gravitation based on Lyra's
geometry. Some of the studies about Lyra theory displayed the
investigation of cosmological models with constant displacement
field vector \cite{{bhamra},{karade},{reddy},{beesham},{soleng}}.
Pradhan and Pandey \cite{pandey} obtained the exact solutions of
bulk viscosity in LRS Bianchi type-I models with constant
deceleration parameter besides Pradhan and Chauhan \cite{pradcha}
obtained the exact solutions of perfect fluid in LRS Bianchi
type-I models with variable deceleration parameter. Rahaman
\emph{et. al.} \cite{kalam} proposed two models according to thin
domain walls in Lyra geometry and pointed out that thin domain
walls have no particle horizons in addition to have gravitational
force, effectively. Rahaman \cite{rahtek} studied global texture
with time dependent displacement vector using weak field
approximation on Lyra geometry. On the other hand, there are some
investigations in homogeneous Friedmann-Robertson-Walker (FRW)
universe with the framework of Lyra geometry in the literature.
For example, Pradhan \emph{et. al.} \cite{prad2} studied bulk
viscosity in FRW-universe and suggested the solutions of energy
density and displacement field vector for power-law or exponential
expansion of the universe in the cases of $k=0$ and $k=-1$. Also,
Singh and Desikan \cite{sindes} displayed the solution of
FRW-universe with time dependent displacement field vector and
constant deceleration parameter, based on Lyra geometry using the
equation of state. Rahaman \emph{et.al.} \cite{rahdas} propounded
the field equations and solutions of higher dimensional
spherically symmetric space-time associated with mass-less scalar field
with constant potential for the flat region.

In this study, we have investigated the models of FRW-universe
with monopoles and domain walls, Chaplygin gas or strange quark
matter in the frame work of the scalar theory of gravitation based on
Lyra geometry. In Section.2, we have obtained the field equations of scalar theory
based on Lyra geometry for given matter groups and universe. We have examined the exact solutions of the field equations for all values of $k=-1,0,1$. Finally, our
results have been discussed.
\cutpage 

\noindent

\section{Field equations of homogeneneous cosmologies in scalar theory based on Lyra geometry}

It is known that FRW-models have worked in varied gravity theories in order to understand nature of universe. A spatially homogeneous and isotropic FRW space-time is

\begin{equation}  \label{eq7}
ds^{2}=dt^{2}-R(t)^2[\frac{1}{1-kr^2}dr^{2}+r^{2}d\theta^{2}+r^{2}sin^{2}\theta d\psi^{2}]
\end{equation}
where curvature parameter, $k$, has values of $k= -1, 0, 1$ according to open,
flat or closed geometry of universe. Also, $R(t)$ is the cosmic scale factor \cite{toplu}.

Moreover, monopoles appeared as point-like defects due to global
symmetry breaking in the evolution of universe according to the standard
cosmology. So, the defects are principally called "global monopoles". Although the
required geometrical conditions for the formation of all defects
are the same, monopoles have more different properties,
physically. Global monopoles hold Lagrangian density as

\begin{equation}  \label{eq8}
\mathcal{L}=\frac{1}{2}\partial_{\mu} \Phi^{i}\partial^{\mu}
\Phi^{i}-\frac{1}{4}\lambda(\Phi^{i}\Phi^{i}-\eta_{m}^2)^2
\end{equation}
where $\Phi^{i}=\eta_{m} f(r)(\frac{x^{i}}{r})$ is a scalar field
of monopoles and $i=1,2,3$. Furthermore, $h(r)$ is equal to zero
at $r=0$ and approaches $h(r)\rightarrow 1$ at $r\gg \delta$,
where the size of monopole core $\delta$ can be described as
$\delta \sim (\sqrt{\lambda} \eta_{m})^{-1}$ \cite{karakitap}.

The energy-momentum tensor of any matter field is widely known as
the form by

\begin{equation}  \label{eq9}
T_{k}^{i}=\partial_{k} \Phi_{l}\partial^{i} \Phi_{l}-\mathcal{L}\delta_{k}^{i}.
\end{equation}
From Eqs.~(\ref{eq7})-(\ref{eq8}) and  Eq.~(\ref{eq9}), the required components
of energy-momentum tensor for global monopoles can be obtained as \cite{karakitap}

\begin{equation}  \label{eq10}
T^{r}_{r}=T^{t}_{t}=\frac{\eta_{m}^2}{r^2}.
\end{equation}

In terms of standart cosmology, domain walls formed due to the
degeneration of discrete symmetry of early universe. In the
Goldstone model, a relation can be identified between domain wall
surface density $\sigma_{w} $ and symmetry breaking scale
$\eta_{w} $ such as $\sigma_{w} \sim \sqrt{\lambda} \eta_{w}^{3}$.
If the symmetry breaking scale is not very
small, it is seeing that domain wall surface density must be dominant from the relation. It
is concluded that domain walls have an enormous role on the
homogeneity of universe \cite{karakitap}. The energy-momentum tensor
of domain walls in the perfect fluid form is given by

\begin{equation}  \label{eq11}
T_{k}^{i}=(p_{m}-
\sigma_{w})(u^{i}u_{k}-\delta_{k}^{i})+(\rho_{m}+\sigma_{w})u^{i}u_{k}
\end{equation}
where $\rho_{m}$ and $p_{m}$ are density and pressure of the matter,
$\sigma_{w} $ is the tension of domain wall, respectively \cite{maeda}.

Recent observations of Type-Ia Supernovae have pointed out that
our universe is spatially flat and has expanded, accelerately
\cite{{perl},{ries},{bachall},{perl2},{bennett},{allen}}. It is believed that the
source of the expansion is the dark energy, which constitutes $\% 70$ of the universe \cite{toplu3}. In the unified model of
dark energy, one of the dark energy candidates is Chaplygin gas,
also called as quartessence \cite{kamen}. The Chaplygin gas has
the similar role with cosmological constant at small or large
values of scale factor in relation with expansion of universe
\cite{{gorini},{alam},{bento}}. In the gravitation theories,
energy-momentum tensor of Chaplygin gas is given by

\begin{equation}  \label{eq12}
T_{k}^{i}=(\frac{A}{\rho_{C}})(u^{i}u_{k}-\delta_{k}^{i})+\rho_{C}u^{i}u_{k}
\end{equation}
where $\rho_{C}$ is density of the Chaplygin gas and $A$ is a negative constant \cite{{kamen},{bento}}.

In the early universe, there are in some other important stages as well
as symmetry breaking. One of them is named Quark-Hadron phase. The
phase in which cosmic temperature had the values of $200 \ MeV$,
passed away from Quark-Gluon Plasma to Hadron gas \cite{ihsan}.
Due to Quark-Hadron phase transition, the quark
matter occurred. According to bag model, using the proportion between the
density and pressure of quark matter such as
$p_{q}=\frac{\rho_{q}}{3}$, total energy density and total
pressure are respectively given by

\begin{equation}  \label{eq13}
\rho_{m}=\rho_{q}+B_{c},
\end{equation}
and

\begin{equation}  \label{eq14}
p_{m}=p_{q}-B_{c}
\end{equation}
where $B_{c}$ is the bag constant \cite{ihsan}. On the other hand,
Equation of State \emph{(EoS)} for strange quark matter is also
given by

\begin{equation}  \label{eq15}
p_{m}=\frac{1}{3}(\rho_{m}-4B_{c}).
\end{equation}
Recently, in the Brookhaven National Laboratory, quark-gluon
plasma has been achieved in the form of perfect fluid, experimentally
\cite{{gorini},{alam},{lu}}. According to the development, it can
be considered Quark-Gluon plasma in the form of perfect fluid and
thus energy-momentum tensor of strange quark matter can be given
by the following form

\begin{equation}  \label{eq16}
T_{k}^{i}=(p_{q}-B_{c})(u^{i}u_{k}-\delta_{k}^{i})+(\rho_{q}+B_{c})u^{i}u_{k}.
\end{equation}

In this study, it can be noted that the matter of space-times has been classified in three different
groups like as (i) monopoles and domain walls, (ii) monopoles and
Chaplygin gas, (iii) monopoles and strange quark matter. Also, it
can be chosen the comoving coordinates as $u^{i}=\delta_{0}^{i}$,
$u^{i}$ stands for the four-velocity. The
displacement vector is used like as $\phi_{i}=(0,0,0,\beta)$ and
$\beta$ is the constant.

\subsection{FRW-cosmologies with monopoles and domain walls in Lyra geometry}
\label{sec:1}
Using the energy momentum tensors of monopoles and domain walls in
Eqs.(\ref{eq10})-(\ref{eq11}) and the line element of FRW
space-time in Eq.(\ref{eq7}) together with Eq.(\ref{eq6}), the
field equations in scalar tensor Lyra theory are obtained the
following form

\begin{equation}  \label{eq17}
\frac{k}{R^2}+\frac{R'^2}{R^2}+\frac{2R''}{R}-\frac{3}{4}\beta^2=-\chi (p_{m}-\sigma_{w}-\frac{\eta_{m}^{2}}{r^2}),
\end{equation}

\begin{equation}  \label{eq18}
\frac{k}{R^2}+\frac{R'^2}{R^2}+\frac{2R''}{R}-\frac{3}{4}\beta^2=-\chi (p_{m}-\sigma_{w}),
\end{equation}

\begin{equation}  \label{eq19}
\frac{3k}{R^2}+\frac{3R'^2}{R^2}+\frac{3}{4}\beta^2=\chi(\rho_{m}+\sigma_{w}+\frac{\eta_{m}^{2}}{r^2}).
\end{equation}
Domain wall density and pressure depend on each other with
\emph{EoS} given by $p_{m}=\gamma \rho_{m}$, where $\gamma$ is the constant. With reference to the \emph{EoS}
and Eqs.(\ref{eq17})-(\ref{eq19}), we have obtained two different exact
solutions of FRW-cosmologies with monopoles and domain walls in
Lyra scalar theory.
\newline

\emph{\textbf{case(i)}} \emph{First solution of FRW-cosmologies
with monopoles and domain walls}
\newline
First set of the exact solutions of FRW-cosmologies with monopoles
and domain walls in Lyra scalar theory has been obtained by
\begin{equation}  \label{eq20}
R(t)=c_{1},
\end{equation}

\begin{equation}  \label{eq21}
\rho_{m}=\frac{1}{\gamma+1}[\frac{2k}{\chi c_{1}^2}+\frac{3}{2\chi}\beta^2],
\end{equation}

\begin{equation}  \label{eq22}
p_{m}=\frac{\gamma}{\gamma+1}[\frac{2k}{\chi c_{1}^2}+\frac{3}{2\chi}\beta^2]
\end{equation}
and

\begin{equation}  \label{eq23}
\sigma_{w}=(\frac{3\gamma+1}{\gamma+1})\frac{k}{\chi c_{1}^2}+(\frac{\gamma-1}{\gamma+1})\frac{3}{4\chi}\beta^2-\frac{\eta_{m}^{2}}{r^2}.
\end{equation}

\bigskip

\emph{\textbf{case(ii)}} \emph{Second solution of
FRW-cosmologies with monopoles and domain walls}
\newline
Second set of the exact solutions of FRW-cosmologies with
monopoles and domain walls in Lyra scalar theory has been obtained
by
\begin{equation}  \label{eq24}
R(t)=\frac{c_{2}}{2}[e^{\frac{(\mp t+c_{3})}{c_{2}}}+k e^{\frac{(\pm t+c_{3})}{c_{2}}}],
\end{equation}

\begin{equation}  \label{eq25}
\rho_{m}=(\frac{1}{\gamma+1})\frac{3}{2\chi}\beta^2,
\end{equation}

\begin{equation}  \label{eq26}
p_{m}=(\frac{\gamma}{\gamma+1})\frac{3}{2\chi}\beta^2
\end{equation}
and

\begin{equation}  \label{eq27}
\sigma_{w}=\frac{3}{\chi c_{2}^2}+(\frac{\gamma-1}{\gamma+1})\frac{3}{4\chi}\beta^2-\frac{\eta_{m}^{2}}{r^2}.
\end{equation}

\subsection{FRW-cosmologies with monopoles and Chaplygin gas in Lyra geometry}
\label{sec:2}
Using the energy momentum tensor of monopoles and Chaplygin gas in
Eqs.(\ref{eq10}), (\ref{eq12}) and the line element of FRW
space-time in Eq.(\ref{eq7}) together with Eq.(\ref{eq6}), the
field equations in scalar tensor Lyra theory are obtained by the
following form

\begin{equation}  \label{eq28}
\frac{k}{R^2}+\frac{R'^2}{R^2}+\frac{2R''}{R}-\frac{3}{4}\beta^2=-\chi(\frac{A}{\rho_{C}}-\frac{\eta_{m}^{2}}{r^2}),
\end{equation}

\begin{equation}  \label{eq29}
\frac{k}{R^2}+\frac{R'^2}{R^2}+\frac{2R''}{R}-\frac{3}{4}\beta^2=-\chi(\frac{A}{\rho_{C}})
\end{equation}

\begin{equation}  \label{eq30}
\frac{3k}{R^2}+\frac{3R'^2}{R^2}+\frac{3}{4}\beta^2=\chi(\rho_{C}+\frac{\eta_{m}^{2}}{r^2}).
\end{equation}
Chaplygin gas density and pressure depend on the each other with
\emph{EoS} given by $p_{C}=\frac{A}{\rho_{C}}$. Starting with the
\emph{EoS} and Eqs.(\ref{eq28})-(\ref{eq30}),
  we have obtained two different exact solutions of FRW-cosmologies with monopoles and Chaplygin gas in Lyra scalar theory.

\bigskip

\emph{\textbf{case(i)}} \emph{First solution of FRW-cosmologies
with monopoles and Chaplygin gas}
\bigskip

\noindent First set of the exact solutions of FRW-cosmologies with
monopoles and Chaplygin gas in Lyra scalar theory has been
obtained by

\begin{equation}  \label{eq31}
R(t)=c_{4},
\end{equation}

\begin{equation}  \label{eq32}
\rho_{C}=-(\frac{k}{\chi c_{4}^2}+\frac{3}{4\chi}\beta^2)\mp\frac{1}{2}\sqrt{(\frac{2k}{\chi c_{4}^2}+\frac{3}{2\chi}\beta^2)^2-4A}
\end{equation}
and

\begin{equation}  \label{eq33}
p_{C}=A[-(\frac{k}{\chi c_{4}^2}+\frac{3}{4\chi}\beta^2)\mp\frac{1}{2}\sqrt{(\frac{2k}{\chi c_{4}^2}+\frac{3}{2\chi}\beta^2)^2-4A} ]^{-1}.
\end{equation}
\bigskip

\emph{\textbf{case(ii)}} \emph{Second solution of
FRW-cosmologies with monopoles and Chaplygin gas}
\bigskip

\noindent Second set of the exact solutions of FRW-cosmologies
with monopoles and Chaplygin gas in Lyra scalar theory has been
obtained by

\begin{equation}  \label{eq34}
R(t)=\frac{c_{5}}{2}[e^{\frac{(\mp t+c_{6})}{c_{5}}}+k e^{\frac{(\pm t+c_{6})}{c_{5}}}],
\end{equation}

\begin{equation}  \label{eq35}
\rho_{C}=-\frac{3}{4\chi}\beta^2\mp\sqrt{\frac{9}{16\chi}\beta^4-A}
\end{equation}
and

\begin{equation}  \label{eq36}
p_{C}=A[-\frac{3}{4\chi}\beta^2\mp\sqrt{\frac{9}{16\chi}\beta^4-A}]^{-1}.
\end{equation}

\subsection{FRW-cosmologies with monopoles and strange quark matter in Lyra geometry}
\label{sec:3}
Using the energy momentum tensors of monopoles and strange quark
matter in Eqs.(\ref{eq10}), (\ref{eq16}) and the line element of
FRW space-time in
 Eq.(\ref{eq7}) with together Eq.(\ref{eq6}), the field equations in scalar
 tensor Lyra theory are obtained by the following form

\begin{equation}  \label{eq37}
\frac{k}{R^2}+\frac{R'^2}{R^2}+\frac{2R''}{R}-\frac{3}{4}\beta^2=-\chi(p_{q}-B_{c}-\frac{\eta_{m}^{2}}{r^2}),
\end{equation}

\begin{equation}  \label{eq38}
\frac{k}{R^2}+\frac{R'^2}{R^2}+\frac{2R''}{R}-\frac{3}{4}\beta^2=-\chi(p_{q}-B_{c})
\end{equation}
and

\begin{equation}  \label{eq39}
\frac{3k}{R^2}+\frac{3R'^2}{R^2}+\frac{3}{4}\beta^2=\chi(\rho_{q}+B_{c}+\frac{\eta_{m}^{2}}{r^2}).
\end{equation}

\noindent Since the strange quark matter has been perfect fluid
form, it's density and pressure depend on each other with
\emph{EoS} given by Eqs.(\ref{eq13})-(\ref{eq15}). From the
Eqs.(\ref{eq13})-(\ref{eq15}) and Eqs.(\ref{eq37})-(\ref{eq39}),
we have two different exact solutions of FRW-cosmologies with
monopoles and strange quark matter in Lyra scalar theory.

\emph{\textbf{case(i)}} \emph{First solution of FRW-cosmologies
with monopoles and strange quark matter}
\newline

\noindent First set of FRW-cosmologies with the exact solutions of
monopoles and strange quark matter in Lyra scalar theory has been
obtained by

\begin{equation}  \label{eq40}
R(t)=c_{7},
\end{equation}

\begin{equation}  \label{eq41}
\rho_{q}=\frac{3}{4}[\frac{2k}{\chi c_{7}^2}+\frac{3}{4\chi}\beta^2],
\end{equation}

\begin{equation}  \label{eq42}
p_{q}=\frac{1}{4}[\frac{2k}{\chi c_{7}^2}+\frac{3}{4\chi}\beta^2]
\end{equation}
and

\begin{equation}  \label{eq43}
B_{c}=\frac{3k}{2\chi c_{7}^2}+\frac{3}{8\chi}\beta^2-\frac{\eta_{m}^{2}}{r^{2}}.
\end{equation}

\emph{\textbf{case(ii)}} \emph{Second solution of
FRW-cosmologies with monopoles and strange quark matter}
\newline
\noindent Second set of the exact solutions of FRW-cosmologies
with monopoles and strange quark matter in Lyra scalar theory has
been obtained by

\begin{equation}  \label{eq44}
R(t)=\frac{c_{8}}{2}[e^{\frac{(\mp t+c_{9})}{c_{8}}}+k e^{\frac{(\pm t+c_{9})}{c_{8}}}],
\end{equation}

\begin{equation}  \label{eq45}
\rho_{q}=\frac{9}{8\chi}\beta^2,
\end{equation}

\begin{equation}  \label{eq46}
p_{q}=\frac{3}{8\chi}\beta^2
\end{equation}
and

\begin{equation}  \label{eq47}
B_{c}=\frac{3}{\chi c_{8}^2}+\frac{3}{8\chi}\beta^2-\frac{\eta_{m}^{2}}{r^2}.
\end{equation}

\section{Conclusion}
Some alternative gravitation theories to Einstein's theory are possible to be seen
in the literature. One of them is accepted as the scalar theory based on Lyra geometry, improved from Weyl theory.
In this study, we have investigated FRW-cosmologies associated with three matter groups,
consisted of monopoles-domain walls, monopoles-Chaplygin gas and monopoles-strange quark matter,
in the framework of Lyra scalar theory of gravitation. The components of displacement vector of
Lyra theory are chosen a constant. It is noted that two exact solutions for each matter groups
are obtained (\emph{see in} \emph{Sec.2}). Our solutions have different
features about cosmic scale factor. In the first solutions, the cosmic scale factor is
constant as given by Eqs.(\ref{eq20}),(\ref{eq31}) and (\ref{eq40}). So, the solutions with constant
$R(t)$, indicate the static universe. In the second solutions, the cosmic scale factor is
time-dependent as given by Eqs.(\ref{eq24}),(\ref{eq34}) and (\ref{eq44}). The solutions have acceleration
and indicate expanding universe. These solutions can be expounded analogously to the following concept.

\begin{figure}
  \includegraphics[width=0.80\textwidth]{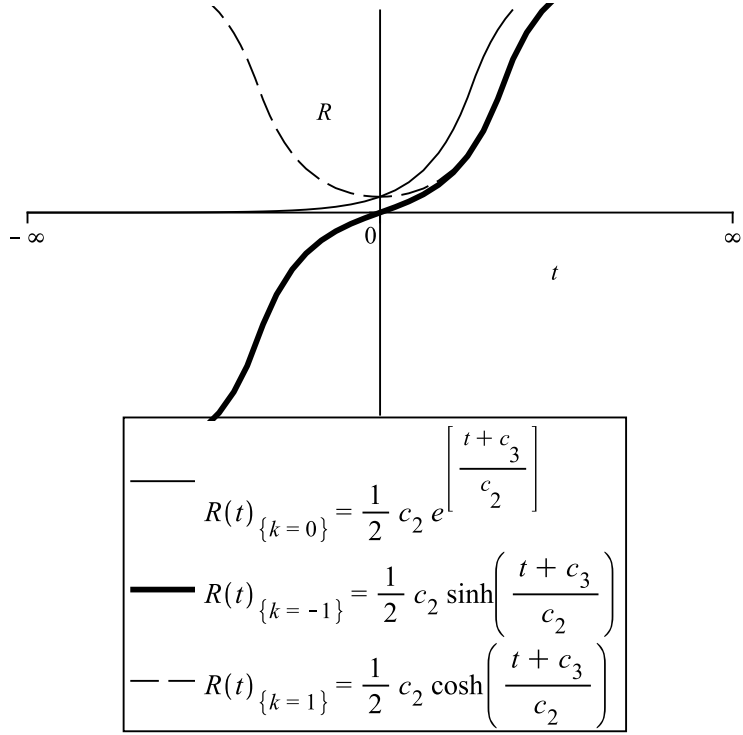}
\caption{The cosmic scale factor in the non-static solution of FRW models with monopole, Chaplygin gas, domain wall or strange quark matter in Lyra geometry}
\label{fig:1}       
\end{figure}

In the second solution of FRW-cosmologies with monopoles
and domain walls, it is clear that the cosmic scale factor $R(t)$
depends on the time, exponentially. The scale factor $R(t)$ in Eq.(\ref{eq24})
transforms the following functions for the values of curvature parameter, $k=0,-1,1$, respectively;

\begin{equation}  \label{eq49}
k\rightarrow 0,\  \  \  \  R(t)=\frac{c_{2}}{2}e^{\frac{\mp t+c_{3}}{c_{2}}},
\end{equation}

\begin{equation}  \label{eq50}
k\rightarrow -1,\  \  \  \ R(t)=\frac{c_{2}}{2}\sinh({\frac{\mp t+c_{3}}{c_{2}}})
\end{equation}
and

\begin{equation}  \label{eq51}
k\rightarrow 1,\  \  \  \ R(t)=\frac{c_{2}}{2}\cosh({\frac{\mp t+c_{3}}{c_{2}}}).
\end{equation}

As the relation with the functions of $R(t)$ in
Eqs.(\ref{eq49})-(\ref{eq51}), it must be called attention to
similar solutions of Friedmann equations for the matter with
pressureless and constant density in Einstein's theory \cite{matts}.  If the models have the expansion and the domain
walls have the non-zero surface tension, the constant $c_{2}$ must
be non-zero ($c_{2}\neq 0$). From Eq.(\ref{eq24}), speed of
expansion and acceleration of universe are given by

\begin{equation}  \label{eq52}
\dot{R}(t)=\frac{1}{2}[\mp e^{\mp\frac{t+c_{3}}{c_{2}}}\pm k e^{\pm\frac{t+c_{3}}{c_{2}}}]
\end{equation}
and

\begin{equation}  \label{eq53}
\ddot{R}(t)=\frac{1}{2c_{2}}[e^{\mp\frac{t+c_{3}}{c_{2}}}+k e^{\pm\frac{t+c_{3}}{c_{2}}}].
\end{equation}

In Einstein's theory, speed and acceleration of universe had the
similar values to Eqs.(\ref{eq52})-(\ref{eq53}) in Lyra theory \cite{matts}.
From the field equation in Eq.(\ref{eq19}) and the solutions in
Eqs.(\ref{eq25}) and (\ref{eq27}), we get

\begin{equation}  \label{eq54}
\frac{\dot{R}(t)^{2}}{R(t)^{2}}=\frac{1}{c_{2}}-\frac{k}{R^{2}}.
\end{equation}

From the Eq.(\ref{eq54}), we get the Hubble parameter/constant as

\begin{equation}  \label{eq55}
k\rightarrow 0, \ \ \ \ H= a,
\end{equation}

\begin{equation}  \label{eq56}
k\rightarrow 1, \ \ \ \ H(t)=[a^{2}-sech^{2}(a(t+c_{3}))]^{\frac{1}{2}}
\end{equation}
and

\begin{equation}  \label{eq57}
k\rightarrow -1, \ \ \ \ H(t)=[a^{2}+cosech^{2}(a(t+c_{3}))]^{\frac{1}{2}}
\end{equation}
where $a=\frac{1}{c_{2}}$. Provided that the field equations of
FRW-cosmologies with monopoles and domain walls in Lyra scalar
theory are compared with the Friedmann equations in de Sitter
cosmology of Einstein's theory, we get the relation between the
Hubble constant $H$ of the solutions in Lyra scalar theory and
cosmological constant $\Lambda$ of Einstein's theory as
\cite{step}

\begin{equation}  \label{eq58}
H=a=\mp\sqrt{\frac{\Lambda}{3}}.
\end{equation}

\begin{figure}
  \includegraphics[width=0.75\textwidth]{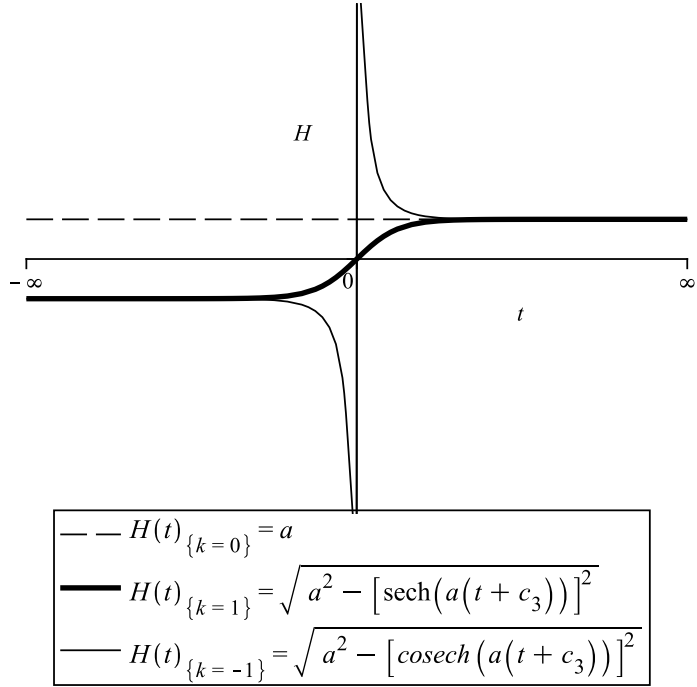}
\caption{Hubble parameter in the non-static solution of FRW models with monopole, Chaplygin gas, domain wall or strange quark matter in Lyra geometry}
\label{fig:2}       
\end{figure}

Also, it is interesting result that surface tension of domain wall
is directly related the subtraction to matter density and pressure
from Eqs.(\ref{eq25})-(\ref{eq27}) as

\begin{equation}  \label{eq59}
\sigma_{w}=\frac{3a^{2}}{\chi}+\frac{(p_{m}-\rho_{m})}{2}.
\end{equation} So, in the flat universe, on condition that $\eta_{m}\rightarrow 0$
and the matter of domain walls is stiffly, ($p_{m}=\rho_{m}$), the surface tension
of domain walls directly depends the Hubble constant as

\begin{equation}  \label{eq60}
\sigma_{w}=\frac{3}{\chi}H^{2}.
\end{equation}
The result has indicated that domain walls can
be responsible for the reason of expansion, in that Hubble
parameter are directly depended the domain wall tension.

It is widely-known that the present universe is expanding with
positive acceleration from observations of Type-Ia Supernovae,
cosmological redshift and Hubble's law \cite{{perl},{ries},{bachall},{perl2},{bennett},{allen}}. It is
suggested that the exotic matter is accounted for the reason of
the expansion. Some candidates of the exotic matter with $p=\gamma
\rho$ ($\gamma<0$), are compiled such as cosmological constant,
dark energy, phantom energy, domain walls, Chaplygin gas and
tachyon. In our solutions, as given by
Eq.~(\ref{eq34}), FRW-cosmologies with Chaplygin gas and
monopoles in the framework of Lyra scalar theory have time-dependent results which grow exponentially and cause an expanding universe, analogously with
FRW-cosmologies with domain walls and monopoles. In as much as
Chaplygin gas and domain walls are already candidates of exotic
matter,
 these solutions in Lyra scalar theory are suitable results, predictably.
 In FRW-cosmologies with Chaplygin gas and monopoles, provided that
$\beta^{4}=\frac{16\chi^{2}}{9}a$ in
Eqs.~(\ref{eq35})-(\ref{eq36}), the matter density and pressure of
Chaplygin gas are equal to others, as $\rho_{C}=p_{C}$. In this
case, the matter of Chaplygin gas is stiff and expanding universe
feature of the model is no degeneration as seen from the
Eq.~(\ref{eq34}).

The FRW-cosmologies with monopoles and strange quark matter in Lyra
scalar theory have the anticipated equation of state,
$p_{q}=\frac{\rho_{q}}{3}$, from Eqs.~(\ref{eq41})-(\ref{eq42})
and (\ref{eq45})-(\ref{eq46}). A result of the models is that Bag
constant of bag model is directly related the subtraction to
matter density and pressure from Eqs.(\ref{eq45})-(\ref{eq47}) as
similar to FRW-cosmologies with domain wall and monopoles, the bag
constant is

\begin{equation}  \label{eq61}
B_{c}=\frac{3}{\chi
c_{8}^{2}}-\frac{(p_{q}-\rho_{q})}{2}-\frac{\eta_{m}^{2}}{r^{2}}.
\end{equation}
 Also, cause of the features of Bag constant, it must be considered the equations near the $r\rightarrow r_{0}$. In the flat universe, on condition that $\eta_{m}\rightarrow
0$, the Bag constant directly depends the Hubble constant as

\begin{equation}  \label{eq62}
B_{c}=\frac{3}{\chi}H^{2}+\frac{\rho_{q}}{3}=-\frac{3}{\chi}H^{2}+p_{q}
\end{equation}
where $H=\frac{1}{c_{8}}$. Because of the perfect fluid form of strange
quark matter, FRW-cosmologies with strange quark matter and
monopoles in Lyra geometry have the solutions which include same features of
expanding universe.

It is sighted that matter density and pressure of domain walls
don't affect from existence of monopoles according to the
Eqs.~(\ref{eq21})-(\ref{eq22}) and Eqs.~(\ref{eq25})-(\ref{eq26}). Otherwise from Eqs.~(\ref{eq17})-(\ref{eq18}), Eqs.~(\ref{eq28})-(\ref{eq29}) and Eqs.~(\ref{eq37})-(\ref{eq38}), it is clearly seen that existence of monopole vanishes cause of $\frac{\eta_{m}^{2}}{r^{2}}=0$. So, in this case, it is called that the scalar tensor theory based on Lyra geometry doesn't allow the solutions of FRW-cosmologies with monopoles. Thus, there are no monopoles in the expanding and accelerated FRW-universe according to the scalar theory based on Lyra geometry.

In this study, for the each of
matter groups, FRW-cosmologies in Lyra scalar theory have two
solutions such as the model of static universe and the model of
time-depended expanding universe. It is possible to correlate
between both of the models. It is emerged that given the first solutions in case(i) of statical universe model is the
particular circumstance of the model of time-depended expanding
universe which is given second solutions in case(ii), provided that $k=0$
at any time ($t=t_{0}$). The case can be pointed out that a part
of time-depended model of universe is similar to take a photograph
of the universe at $t=t_{0}$, instantaneously. In contrast, first
solutions completely wide apart from second solutions, provided
$k\neq 0$. The obtained solutions of time-depended expanding model
for FRW-cosmologies with monopoles and domain walls, Chaplygin gas
or strange quark matter in Lyra scalar theory agree with the
solutions for FRW-cosmologies with perfect fluid in Einstein's
theory. But, it is obtained the completely different solutions
called static model in the Lyra scalar theory. Halford \cite{halford} proposed that constant displacement vector $\Phi_{i}$, therefore number $\beta^{2}$ in Lyra scalar theory of gravitation, play the role of the cosmological constant $\Lambda$ in Einstein's theory.
Halford's suggestion has been extensively informed in many
studies, in the literature \cite{{rahgos},{sindes},{hal1},{hal2},{hal3},{hal4},{hal5},{hal6},{hal7},{hal8},{hal9},{hal10},{hal11},{hal12},{hal13},{hal14},{hal15},{hal16}}. Even if, Halford can be
right in his conjecture, his suggestion must be modified due to contain
inadequate information. In the case of constant displacement
vector like as $\beta=constant$, it can be obtained solutions of field equations in Lyra scalar
theory. These solutions can agree with the solutions which is obtained solutions of field equations in Einstein's
theory. But, Lyra scalar theory has the more general solutions
than Einstein's theory. Because, in field equations, there are two terms of
$\Phi_{i}\Phi^{k}$ and $\Phi_{m}\Phi^{m}g_{ik}$ in Lyra scalar
theory besides the term with cosmological constant of $\Lambda
g_{ik}$ in Einstein's theory. Provided to $\beta=constant$, the
term of $\Phi_{m}\Phi^{m}g_{ik}$ in Lyra scalar theory can be
equivalent with the term of $\Lambda g_{ik}$ in Einstein's theory.
But it is clearly that the term of $\Phi_{i}\Phi^{k}$ in Lyra
scalar theory can add a constant term to field equations unlike in
Einstein's theory. Because the term of $\Phi_{i}\Phi^{k}$ in
Lyra scalar theory does not multiply the metric potential
$g_{ik}$, the field equations have the extra term in comparison with Einstein's theory. Thus, the field
equations in Lyra theory must be different from fields equations
in Einstein's theory. This means that obtained solutions from field
equations in Lyra theory must be more general from Einstein's, even though displacement vector has constant components.

Also, it can be seen from the solutions in this study, in addition to
this result which can be directly emerged with to compare field
equations of both theories. Firstly, we get the model of static
universe for the FRW-cosologies in Lyra scalar theory. In the
Einstein's theory, there is no similar to the solution for
Friedmann equations. On the other hand, the Hubble parameter for
FRW-cosmologies in the Einstein's theory depends on the
cosmological constant as Eq.~(\ref{eq58}). Already, the
cosmological constant is considered to be responsible for the
expansion of universe. If widely-known consider in literature could be correct
completely, the reason of the expansion of universe would need to be the
constant of $\beta$ which called to play the role of cosmological constant,
in Lyra scalar theory. So, Hubble parameter in Lyra scalar theory
would need to depend on the constant of $\beta$, but it is seen that the
Hubble parameter in Lyra scalar theory doesn't depend on the
constant of $\beta$ in view of Eqs.~(\ref{eq55})-(\ref{eq57}). For
example, the Hubble parameter in flat universe is the constant which is different from $\beta$,
getting arbitrarily from obtained the solutions.

Consequently, the suggestions of many authors must be generalized:
"field equations in Einstein theory could be the particular situation of field equations in Lyra scalar
theory, provided that $\beta=constant$. Lyra scalar theory is the
equivalent theory of Einstein with no cosmological constant
provided that $\beta=0$, completely."

\section{Acknowledgements}
\label{sec:4}
This work is supported by Scientific Research Project of \c{C}anakkale Onsekiz Mart University under the Grant $2012/006$.

\bibliographystyle{my-h-elsevier}

\begin{thebibliography}{10}






\bibitem{lyra} G. Lyra, \"{U}ber eine modifikation der rieamannschen geometrik, Math. Z {\bf 54} (1951) 52.
\bibitem{brans} C. Brans and R. H. Dicke, Mach's principle and a relativistic theory of gravitation, Phys. Rev. {\bf 124} (1961) 925.
\bibitem{rahgos} F. Rahaman and P. Ghosh, Gravitational field of domain wall in Lyra geometry, Astrophys. Space Sci. {\bf 317} (2008) 127-132.
\bibitem{weyl} H. Weyl, Vorlesungen \"{u}ker Allgemeine relativit\"{o}tstheorie, Sber. Preuss. Akad. der Wiss. {\bf 465} Sitzunbgsberichte, zu Berlin (1918).
\bibitem{pradplane} A. Pradhan, V. Rai and S. Otarod, Plane symmetric inhomogeneous bulk viscous domain wall in Lyra geometry, Fizika B {\bf 15} (2006) 57-70.
\bibitem{rahmalgos} F. Rahaman, S. Mal and P. Ghosh, A study of global monopole in Lyra geometry, Mod. Phys. Lett. A {\bf 19} (2004) 2785-2790.
\bibitem{pradcha} A. Pradhan and D. S. Chauhan, A new class of LRS Bianchi Type-I cosmological model in Lyra geometry, 	RAPC {\bf 8} (2006) 179-190.
\bibitem{kitap} V. Faraoni, Cosmology in scalar-tensor gravity, Fundamental theories of physics. Kluwer Academic Publishers, Netherland, (2004).
\bibitem{sen} D. K. Sen, A static cosmological model, Z.f\"{u}r Physik {\bf 149} (1957) 311.
\bibitem{sendunn} D. K. Sen and K. A. Dunn, A scalar-tensor theory of gravitation in a modified Riemannian manifold, J. Math. Phys. {\bf 12} (1971) 578.
\bibitem{halford} W. D. Halford, Cosmological theory based on Lyra's geometry, Austr. J. Phys. {\bf 23} (1970) 863.
\bibitem{bhamra} K. S. Bhamra, A cosmological model of class one on Lyra's manifold, Austr. J. Phys., {\bf 27} (1935) 541.
\bibitem{karade} T. M. Karade and S. M. Borikar, Thermodynamic equilibrium of a gravitating sphere in Lyra geometry, Gen. Rel. Grav. {\bf 9} (1978) 431.
\bibitem{reddy} D. R. K. Reddy and P. Innaiah, A plane symmetric cosmological model in Lyra manifold, Astrophys. Space Sci. {\bf 49} (1986) 123.
\bibitem{beesham} A. Beesham, Vacuum Friedmann cosmology based on Lyra's manifold, Astrophys. Space Sci. {\bf 127} (1986) 189.
\bibitem{soleng} H. H. Soleng, Cosmologies based on Lyra's geometry, Gen. Rel. Grav. {\bf 19} (1987) 1213.
\bibitem{pandey} A. Pradhan and H. R. Pandey, Bulk viscous cosmological models in Lyra geometry, Int. J. Mod. Phys. D {\bf 10} (2001) 3.
\bibitem{kalam} F. Rahaman, M. Kalam and R. Mondal, Thin domain walls in Lyra geometry, Astrophys. Space Sci. {\bf 305} (2006) 337.
\bibitem{rahtek} F. Rahaman, Global texture in Lyra geometry, NuovoCim B {\bf 118} (2003) 17.
\bibitem{prad2} A. Pradhan, V. K. Yadav, I. Chakrabarty and D. V. Ahluwalia, Bulk viscous FRW cosmology in Lyra geometry, Int. J. Mod. Phys. D {\bf 10 (03)} (2001) 339-349.
\bibitem{sindes} G. P. Singh and K. Desikan, A new class of cosmological models in Lyra geometry, Pramana {\bf 49 (2)} (1997) 205-212.
\bibitem{rahdas} F. Rahaman, N. Begum and S. Das, A class of higher dimensional spherically symmetric cosmological model in Lyra geometry, Astro. Phys. Space. Sci. {\bf 249 (3-4)} (2004) 219-224.
\bibitem{toplu} A. Friedmann, \"{U}ber die M\"{o}glichkeit einer Weit mit konstanter negativer Kr\"{u}mmung des Raunes, Zeitschrift für Physik A {\bf 21 (1)} (1924) 326-332; G.Lemaitre, L'Univers en expansion, Annales de la Soci\'{e}t\'{e} Scientifique de Bruxelles A {\bf 53} (1933) 51-56; H. P. Robertson, Kinematics and world-structure, Astrophys. J. {\bf 82} (1935) 284; A. G. Walker, Proceedings of the London Mathematical Society 2 {\bf 42 (1)} (1937) 90-127.
\bibitem{karakitap} A. Vilenkin and E.P.S. Shellard, Cosmic strings and other topological defects, Cambridge University Press, (1994).
\bibitem{maeda} N. Okuyama and K. Maeda, Domain wall dynamics in brane world and nonsingular cosmological models, Phys. Rev. D {\bf 70} (2004) 08405.
\bibitem{perl} S. Perlmutter, G. Aldering, S. Deustua, S. Fabbro, G. Goldhaber, D. E. Groom, A. G. Kim, R. A. Knop, P. Nugent, C. R. Pennypacker, M. della Valle, R. S. Ellis, R. G. McMahon, N. Walton, A. Fruchter, N. Panagia, A. Goobar, I. M. Hook, C. Lidman, R. Pain, P. Ruiz-Lapuente, B. Schaefer and Supernova Cosmology Project, Cosmology from Thype IA Supernovae:Measurements, calibration techniques and implications, Bull. Am. Astron. Soc. {\bf 29} (1997) 1351.
\bibitem{ries} A. G. Riess, A. V. Filippenko, P. Challis, A. Clocehiatti, A. Diercks, P. M. Garnavich, R. L. Gilliland, C. J. Hogan, S. Jha, R. P. Kirshner, B. Leibundgut, M. M. Phillips, D. Reiss, B. P. Schmidt, R. A. Schommer, R. C. Smith, J. Spyromilio, C. Stubbs, N. B. Suntzeff and J. Tonry, Observational evidence from Supernovae for an accelerating universe and a cosmological constant, Astron. J. {\bf 116} (1998) 1009-1038.
\bibitem{bachall} N. A. Bachall, J. P. Ostriker, S. Perlmutter and P. J. Steinhardt, The cosmic triangle: Revealing the state of the universe, Science {\bf 284} (1999) 1481.
\bibitem{perl2} S. J. Perlmutter, G. Aldering, G. Goldhaber, R. A. Knop, P. Nugent, P. G. Castro, S. Deustua, S. Fabbro, A. Goobar, D. E. Groom, I. M. Hook, A. G. Kim, M.Y. Kim, J. C. Lee, N. J. Nunes, R. Pain, C.R. Pennypacker, R. Quimby, C. Lidman, R. S. Ellis, M. Irwin, R. G. McMahon, P. Ruiz-Lapuente, N. Walton, B. Schaefer, B. J. Boyle, A. V. Pilippenko, T. Matheson, A. S. Fnuchter, N. Panagia, H. J. M. Newberg, W. J. Couch and Supernova Cosmology Project, Measurements of Omega and Lambda from 42 high-redshif Supernovae, Astrophys. J. {\bf 517} (1999) 565-586.
\bibitem{bennett} C. L. Bennett, M. Halpern, G. Hinshaw, N. Jarosik, A. Kogut, M. Limon, S. S. Meyer, L. Page, D. N. Spergel, G. S. Tucker, E. Wollack, E. L, Wright, C. Barnes, M. R. Greason, R. S. Hill, E. Komatsu, M. R. Nolta, N. Odegard, H. V. Peiris, L. Verde and J.L. Weiland, First year Wilkinson Microwave Anizotropy Probe (WMAP) observations preliminary maps and basic results, Astrophys. J. Suppl. {\bf 148} (2003) 1-27.
\bibitem{allen} S. W. Allen, R. W. Schmidt, H. Ebeling, A. C. Fabian and L. van Speybroeck, Constraints on dark energy from Chandra observations of the largest relaxed galaxy clusters, Mon. Not. Roy. Astron. Soc. {\bf 353} (2004) 457-467.
\bibitem{toplu3} V. Sahni and A. A. Starobinsky, The case for a positive cosmological $\Lambda$-term, Int. J. Mod. Phys. A {\bf 9} (2000) 373-443; P. J. E. Peebles and B. Ratra, The cosmological constant and dark energy, Rev. Mod. Phys. {\bf 75} (2003) 559-606; T. Padmanabhan, Cosmological constant- the weight of the vacuum, Phys. Rept. {\bf 380} (2003) 235-320.
\bibitem{kamen} A. Kamenshchik, U. Moschella and V. Pasquier, An alternative to quintessence, Phys. Lett. B {\bf 511} (2001) 265-268.
\bibitem{gorini} V. Gorini, A. Kamenshchik and U. Moschella, Can the Chaplygin gas be a plausible model for dark energy?, Phys. Rev. D {\bf 67} (2003) 063509.
\bibitem{alam} U. Alam, V. Sahni, T. Deep Saini and A. A. Starobinsky, Exploring the expanding Universe and dark energy using the statefinder diagnostic, Mon. Not. R. Astron. Soc. {\bf 344} (2003) 1057-1074.
\bibitem{bento} M. C. Bento, O. Bertolami and A. A. Sen, Generalized Chaplygin gas accelerated expansion and dark-energy-matter unification, Phys. Rev. D {\bf 66} (2002) 043507.
\bibitem{ihsan} \.{I}. Y{\i}lmaz, String cloud and domain walls with quark matter in 5-D Kaluza Klein cosmological model, Gen. Rel. Grav. {\bf 38} (2006) 1397-1406.
\bibitem{lu} H. Q. Lu, Z. G. Huang, W. Fang and K. F. Zhang, Dark Energy and Dilaton Cosmology, JCAP {\bf 03} (2008) 12.
\bibitem{matts} M. Roos, Introduction to cosmology, Second edition, John Wiley and Sons Ltd., England (1994).
\bibitem{step} H. Stephani, General Relativity, Cambridge University Press (1982).
\bibitem{hal1} A. Pradhan, I. Aotemshi and G. P. Singh, Plane Symmetric Domain Wall in Lyra Geometry, Astrophys.Space Sci. {\bf 288} (2003) 315-325.
\bibitem{hal2} F. Rahaman, S. Das, N. Begum and M. Hossain, Higher Dimensional Homogeneous Cosmology in Lyra Geometry, Pramana {\bf 61 (1)} (2003) 153-159.
\bibitem{hal3} F. Rahaman, S. Chakraborty, N. Begum, M. Hossain and M. Kalam, Bianchi-IX String Cosmological Model in Lyra Geometry, Pramana {\bf 60 (6)} (2003) 1153-1159.
\bibitem{hal4} F. Rahaman, S. Chakraborty, S.Das, M. Hossain and J. Bera, Higher Dimensional String Theory in Lyra Geometry, Pramana {\bf 60 (3)} (2003) 453-459.
\bibitem{hal5} B. B. Bhowmik and A. Rajput, Anisotropic Lyra Cosmology, Pramana Jour. of Phys. {\bf 62 (6)} (2004) 1187-1199.
\bibitem{hal6} F. Rahaman, K. Maity, P. Gosh and K. Gayen, Gravitational Field of Higher Dimensional Global Monopole in Lyra Geometry, Bulletin of the Gauhati University Mathematics Association {\bf 8} (2004) 82.
\bibitem{hal7} F. Rahaman, Sk. M.  Hossain, S. Mal, P. Gosh and R. Mukherji, A Study of Global String in Lyra Geometry, Fizika B {\bf 14 (1)} (2005) 303.
\bibitem{hal8} F. Rahaman, S. Mal, S. Shekhar, P.Gosh and T. Ray, Semi-Classical Gravitational Effects Near Cosmic String in Lyra Geometry, Fizika B {\bf 14 (1)} (2005) 327.
\bibitem{hal9} F. Rahaman and R. Mondal, Non Static Global Monopole in Lyra Geometry, Fizika B {\bf 16 (1)} (2007) 223.
\bibitem{hal10} A. Pradhan and P. Mathur, Inhomogeneous Perfect Fluid Universe with Electromagnetic Field in Lyra Geometry, Fizika B {\bf 18} (2009) 243-264.
\bibitem{hal11} R. M. Gad, Axially Symmetric Cosmological Mesonic Stiff Fluid Models in Lyra's Geometry, arXiv:0909.1503v2 (2009).
\bibitem{hal12} A. K. Yadav, Lyra's Cosmology of Inhomogeneous Universe with Electromagnetic Field, Fizika B {\bf 19} (2010) 53-80.
\bibitem{hal13} S. Agarwal, R. K. Pandey and A. Pradhan, Bianchi Type-II String Cosmological Models in Normal Gauge for Lyra's Manifold with Constant Deceleration, arXiv:1010.1947v1 (2010).
\bibitem{hal14} A. K. Yadav and A. Haque, Lyra's Cosmology of Massive String in Anisotropic Bianchi-II Sace-time, Int. J. Theor. Phys. {\bf 50} (2011) 2850-2863.
\bibitem{hal15} S. Agarwal, R. K. Pandey and A. Pradhan, LRS Bianchi Type II Perfect Fluid Cosmological Models in Normal Gauge for Lyra's Manifold, Int. J. Theor. Phys. {\bf 50} (2011) 296-307.
\bibitem{hal16} A. Pradhan, H. Amirhashchi and H. Zainuddin, A New Class of Inhomogeneous Cosmological Models with Electromagnetic Field in Normal Gauge for Lyra's Manifold, Int. J. Theor. Phys. {\bf 50} (2011) 56-69.







\end{thebibliography}

\end{document}